\documentclass[proof]{WileyASNA-v1}

\articletype{Article Type}%

\received{1 October 2020}
\revised{12 October 2020}
\accepted{25 October 2020}

\begin{document}

\title{The binding energy produced within the framework of the accretion of millisecond pulsars}

\author[1]{Ali Taani}


\authormark{Ali Taani}

\address[1]{\orgdiv{Physics Department, Faculty of Science, Al-Balqa Applied University, 19117 Salt, Jordan}}


\corres{Ali Taani.  \email{ali.taani@bau.edu.jo}}
\abstract{The role and implication of binding energy through the accretion-induced collapse (AIC) of accreting white dwarfs (WDs) for the production of millisecond pulsars (MSPs) are  investigated. I examine the binding energy model due to the dynamical process in close binary systems and investigating the possible mass of the companion sufficient to induce their orbital parameters. The deterministic nature of this interaction has a strong sensitivity to the equation of state of the binary systems (where the compactness of a neutron star is proportional to the amount of binding energy) associated with their initial conditions. This behavior will mimic the commonly assumed mass and amount of accreted matter under the instantaneous mass loss ($\Delta M \sim 0.18M_{\odot}$). As a result, this will indicate an increase in the MSP's gravitational mass due to angular momentum losses. The outcome of such a system will then be a circular binary MSP in which the companion is a low-mass WD, thus distinguishing the binary formation scenarios. In addition, the results of this work could provide constraints on the expected mass and binding energy of a neutron star based on the accretion rate.}


\keywords{binding energy, pulsars, evolution, kinematics and dynamics.}

\maketitle

\section{Introduction}

Millisecond pulsars (MSPs) are remarkable objects born via type II supernovae (SNe)
explosions and characterized by short spin periods (P$_{spin} \leq $ 20 ms), weak magnetic fields (B $\leq 10^{9}$ G) extremely old
 $ \sim 10^{9}$ yr based on recycling process. ATNF census of Galactic 445 MSPs in both the Galactic
plane and in globular clusters 
\citep[see e.g.,][]{2005AJ....129.1993M, 2014Freire, 2012ChA&A..36..137C}.  They are frequently found in binaries (about 65\%) with white dwarfs (WDs) in circular orbits with companions with masses ranging from $\rm \sim 0.15 M_{\odot} -  0.45M_{\odot}$ \citep[see e.g.,][]{1982Natur.300..728A, 1991PhR...203....1B, 2012AN....333...53T,  2017JPhCS.869a2090T}.

 \begin{figure}[htb]
\centering
\includegraphics[width=80mm,height=90mm]{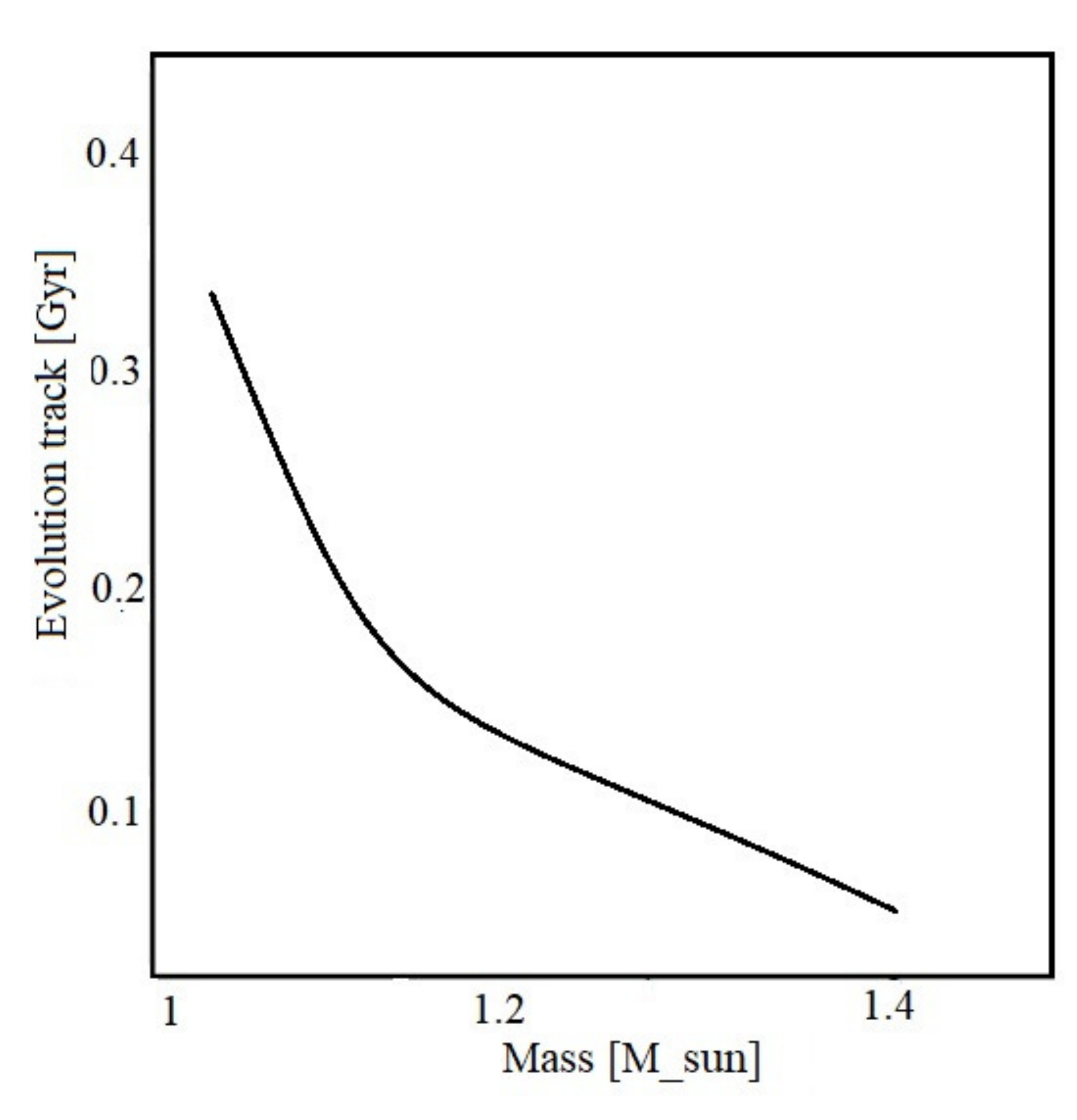}
\caption{The AIC of binary system evolution is tracked, and the  mass accretion-rate is expected to be around ($\Delta M \rm \sim 0.2M_{\odot}$) with a  timescale shorter than 0.4 Gyr \citep[see e.g.,][]{2002MNRAS.329..897H, 2013IJMPS..23...95J}.} \label{fig1}
\end{figure}

  \begin{figure}[htb]
\centering
\includegraphics[width=80mm,height=90mm]{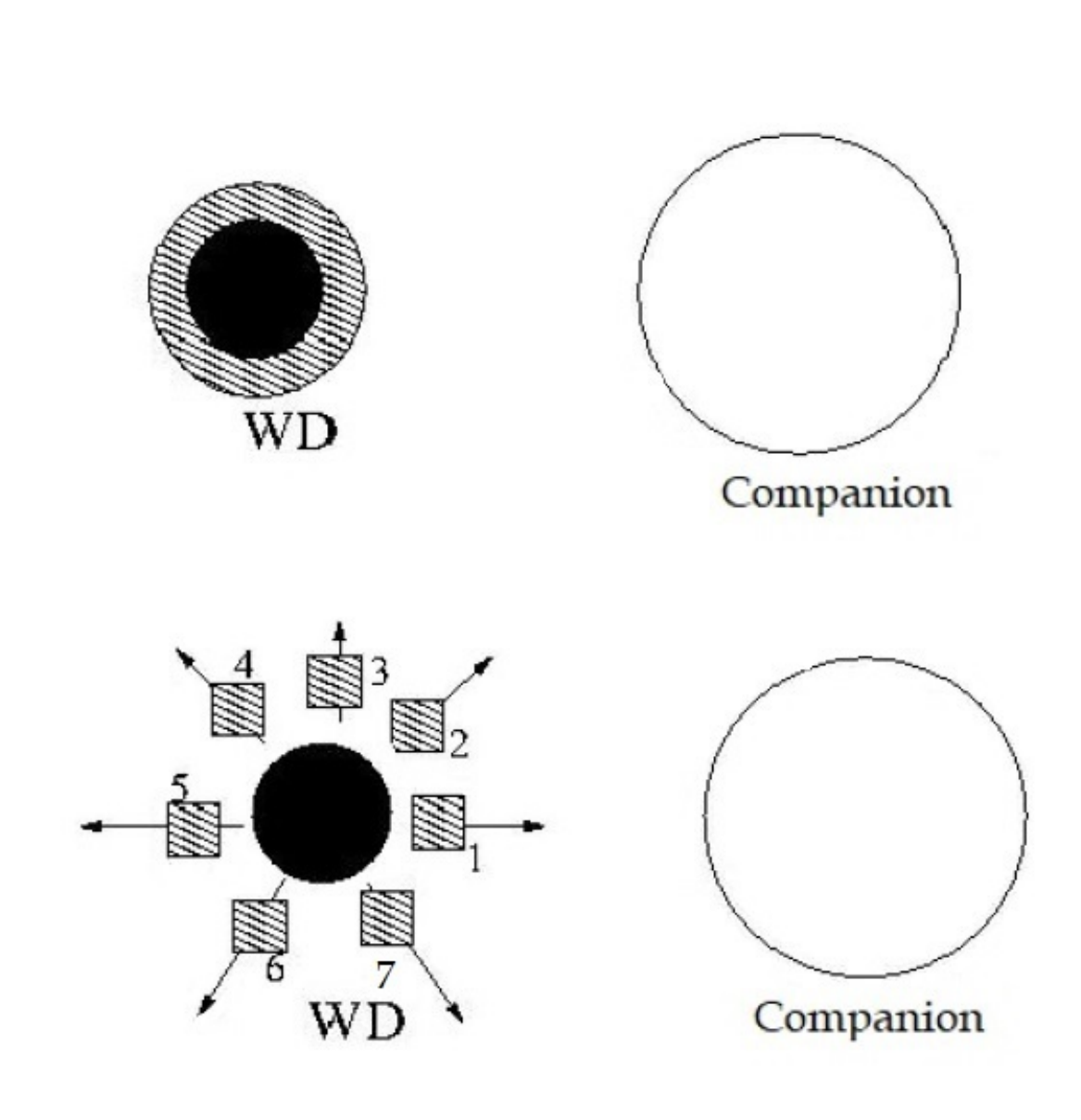}
\caption{ Simple diagram illustrating how the companion star, $M_c$, will receive some of the mass ejected by the exploding WD during the AIC. The top panel corresponds to an instant right before the explosion. The bottom panel corresponds to an instant right after it happened. The ejected material has been schematically divided into blobs 1, 2, 3, 4, 5, 6, 7.  The sum of the masses of all blobs is asymmetric mass ejection. The motion of each blob has been shown with an arrow. The blobs 1, 2 and 7 will reach the surface of $M_c$. Therefore, the mass of blobs 1, 2 and 7 are  received and set to be ($\Delta M \rm \sim 0.18M_{\odot}$).} \label{fig2}
\end{figure}

It was suggested that MSPs were formed in low-mass X-ray
binary systems (LMXBs).  This argument called for a recycling
process in which the slowly rotating, old Neutron Star (NS) may be spun-up into an MSP via accretion from a binary companion  \citep[see e.g.,][]{1982Natur.300..728A, 1991PhR...203....1B, 2006Tauris, 2019AN....340..847T, 2017PASA...34...24T}.
This link has been more recently reinforced by the discovery of many pulsars like PSR J1023+0038 \citep{2009Sci...324.1411A}, whose radio emission is likely to have  switched on only recently after the LMXB phase.

An alternative formation scenario is the accretion induced collapse (AIC) of
a ONeMg white dwarf (WD) \citep[see e.g.,][]{1987ApJ...322..206N, 2010MNRAS.402.1437H, 2013Tauris, 2014Freire, 2017JPhCS.869a2090T, 2017ApJ...846..170T, 2019Schwab, 2020RAA....20..135W}. In this scenario, the mass transfer can
increase the mass of a WD toward the Chandrasekhar mass. When the star reaches this mass limit, degeneracy pressure can no longer support it because the supporting electron pressure is robbed by inverse $\beta$
decays \citep{2005ASPC..334...65K}, consequently  losing enough energy in
neutrinos.  This causes it to collapse and violently release a substantial
amount of gravitational energy, which might be observable by the gravitational wave observatories such as LIGO-Virgo
Collaboration and GEO \citep{2009MNRAS.396.1659M, 2010MNRAS.409..846D, Mardini_2019a, Mardini_2019b, Mardini_2019c, Mardini2020, 2022MNRAS.510.6011W}.

The NS equation of state, along with the mass-radius relation, plays a role in modeling the baryonic and gravitational masses as well as binding energy \citep{2009PhRvC..80f5809N}. The majority of efforts have been made to calculate the effect of the binding energy by estimating the amount of evolution driven by mass accretion in order to understand the evolutionary tracks of binary models \citep[see e.g.,][]{2011MNRAS.413L..47B,  2018MNRAS.477.2349I, 2019RAA....19...12T, 2019JPhCS1258a2029T, 2021arXiv210313605H, 2021ApJ...910L..22H, 2004ApJ...614..914A}.

\cite{2021arXiv210313605H} constrained  the baryonic mass and binding energy associated with the equation of state, of two double pulsars (J0737-3039B and J1756-2251), finding that  are thought to have formed from an ultra-stripped progenitor. Also he found that these systems  are consistent with forming from the AIC.

This work aims  to consider the pre-AIC and post-AIC binding energies
of the binary system. Then, taking into account the non-zero velocity of ejected matter, I derive several elementary analytic formulae for evaluating the variation of orbital energy and the possibility of binary survival.

 \begin{figure}[htb]
\centering
\includegraphics[width=80mm,height=90mm]{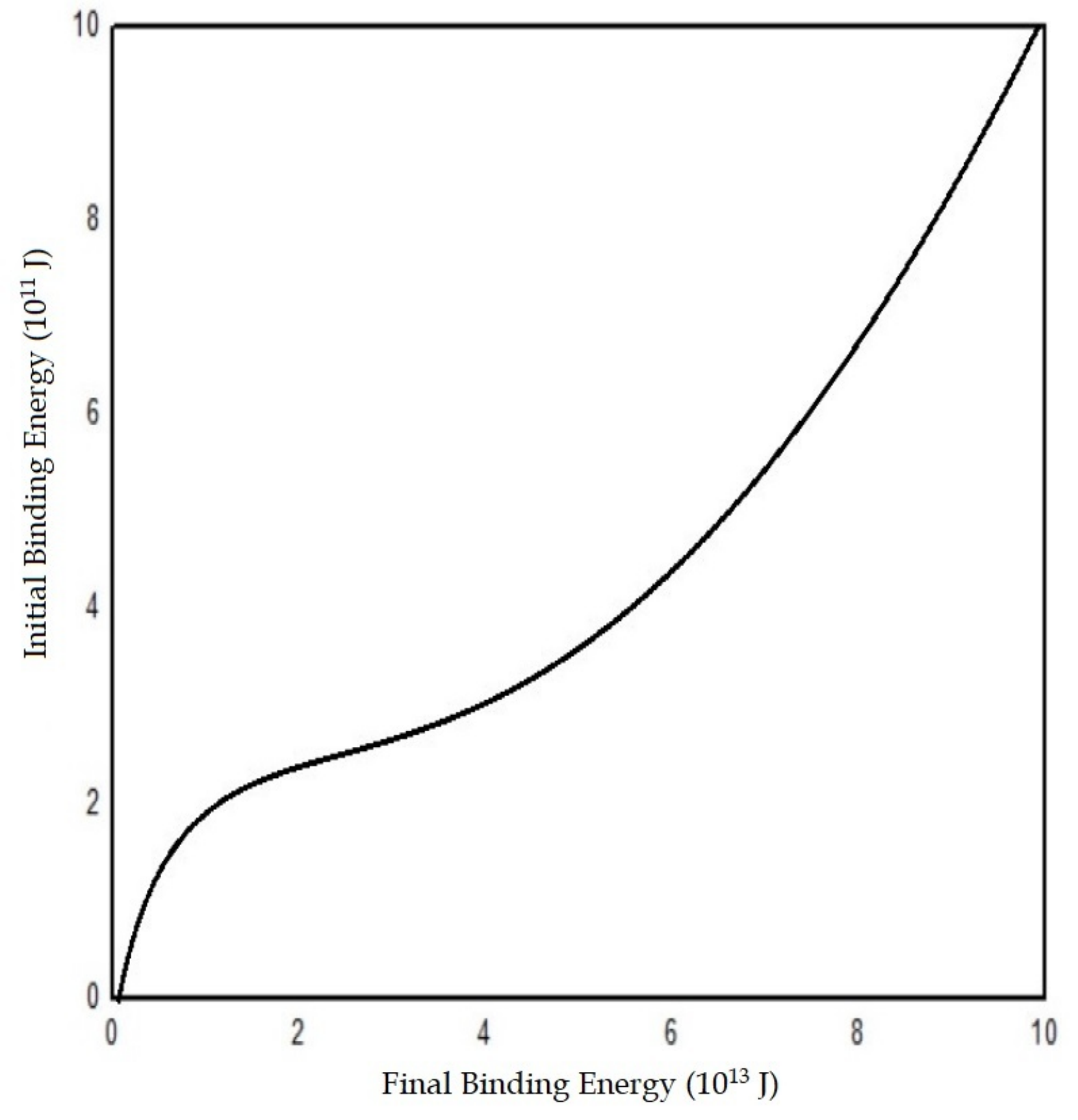}
\caption{The dynamical evolution of the post-AIC and pre-AIC systems as we can be seen the initial and final binding energies. Note that the  typical mass loss  taken for this is   $0.18M_{\odot}$.} \label{fig3}
\end{figure}

 \begin{figure}[htb]
\centering
\includegraphics[width=80mm,height=90mm]{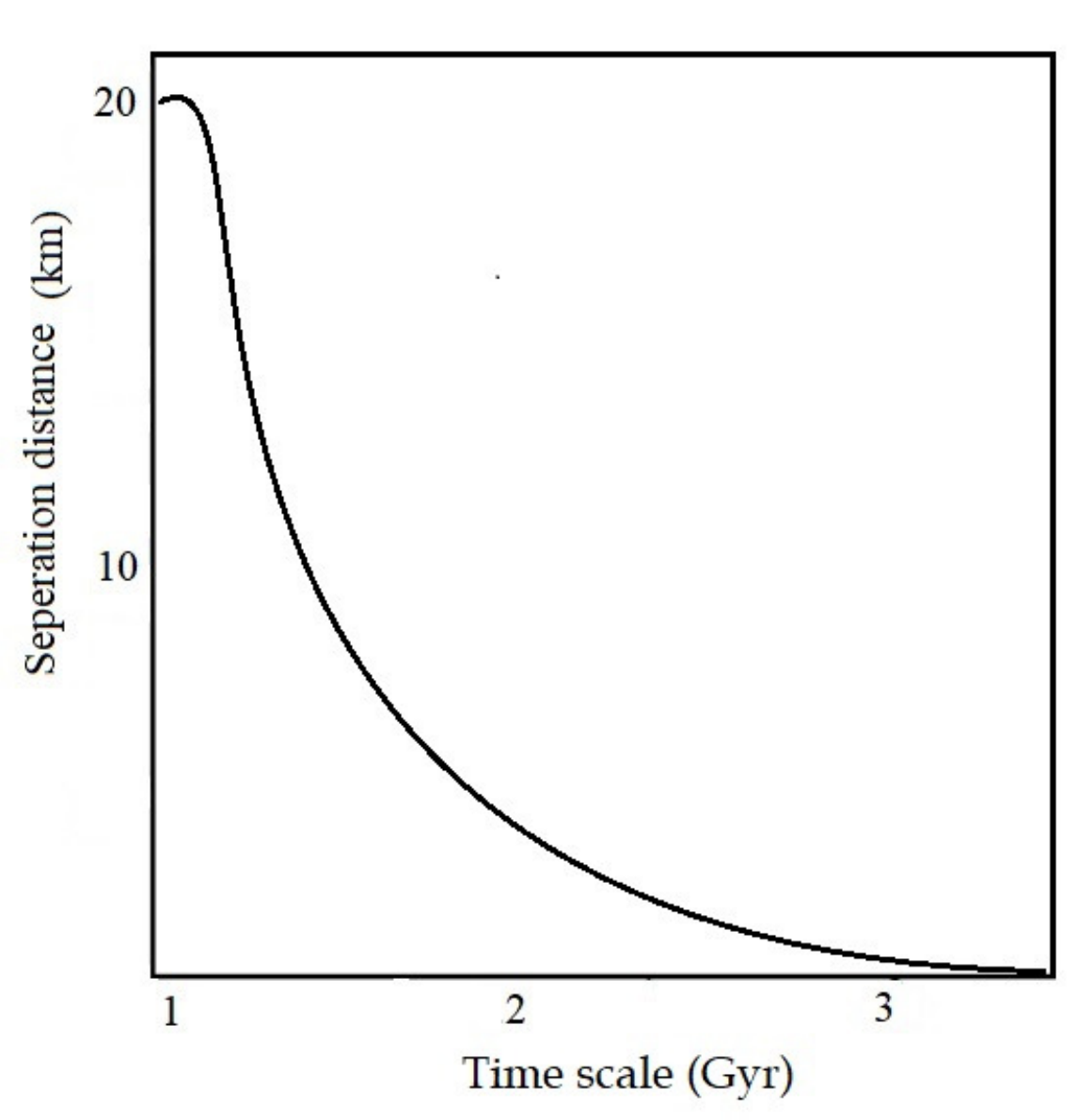}
\caption{The final separation distance as a function of time. An indicator of the presence of a dynamical AIC process  in a given system is the  strong sensitivity to initial conditions i. e. masses and semi-major axes (see system details in the text).} \label{fig4}
\end{figure}

\begin{figure}[htb]
\centering
\includegraphics[width=80mm,height=90mm]{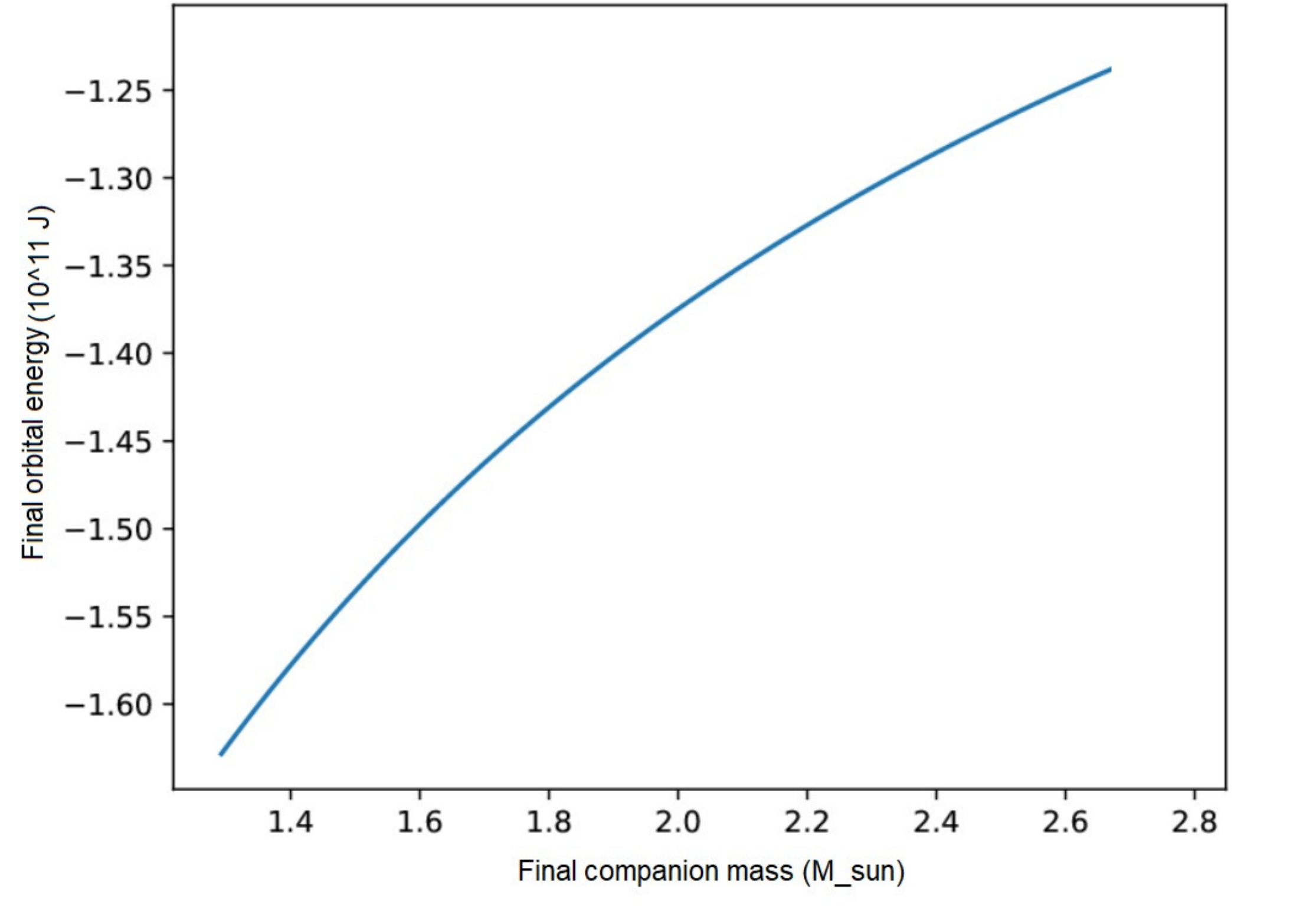}
\caption{The final orbital energy as a function of the final fate of the mass companion. The correlation is  roughly linear, as predicted by the H-R diagram.} \label{fig5}
\end{figure}

\begin{figure}[htb]
\centering
\includegraphics[width=80mm,height=90mm]{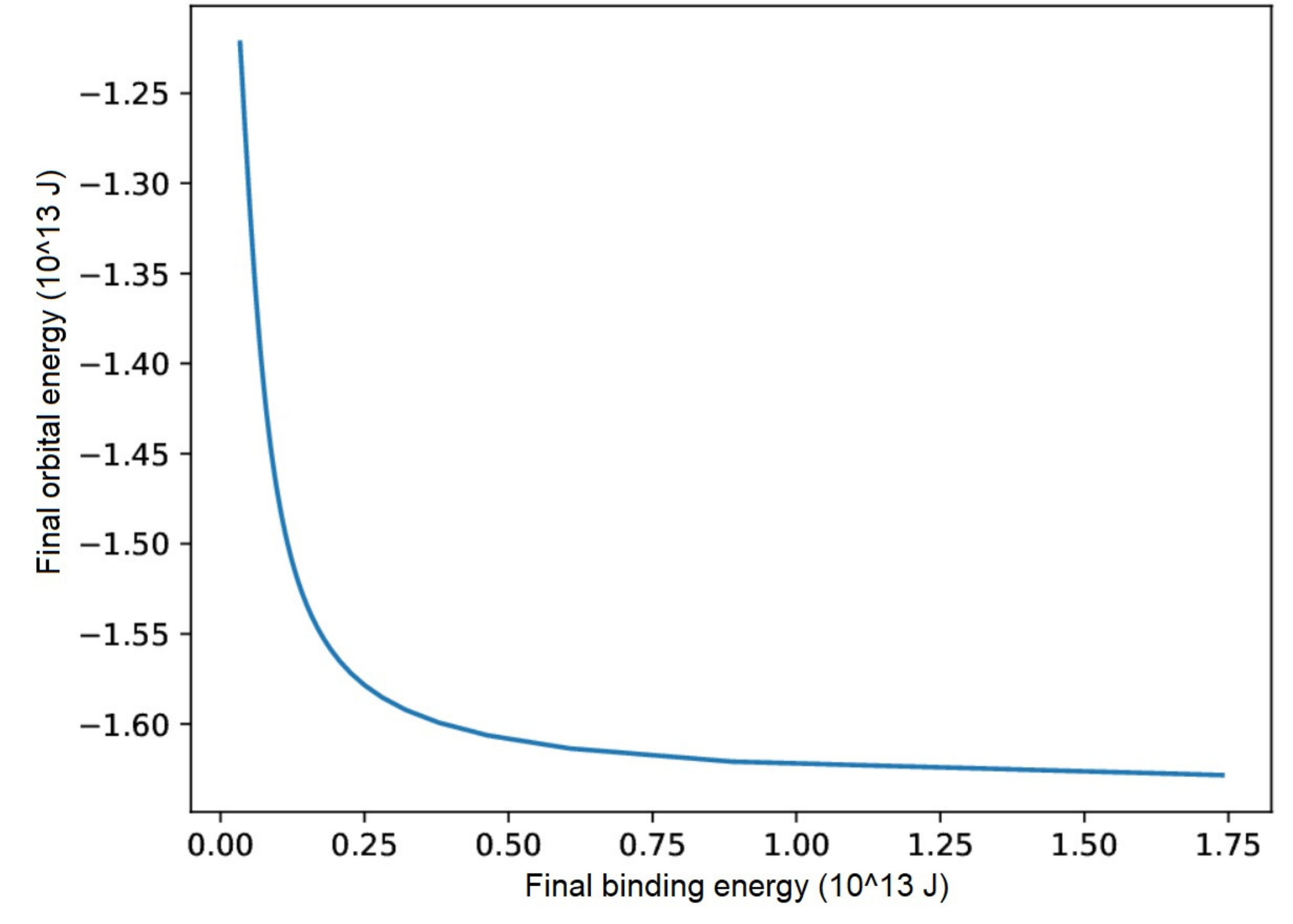}
\caption{The final orbital energy as a function of the final binding energy through the AIC process.} \label{fig6}
\end{figure}

\section{The Binding Energy}
The impact of kick velocity on the companion depends on the binding energy
of the system and  the kinetic energy of the kick exerted on the orbit. 
It is noteworthy to mention here that, after AIC, the mass transfer will start once more due to the evolution of the companion star. Fig. ~\ref{fig1}, shows the evolutionary track of the AIC process as it decreases  mass. However,  given a particular initial mass of the WD $1.1M_{\odot}$ (i.e.  Taani et al. 2012a), it will take a time interval shorter than 0.4 Gyr \citep[see e.g.,][]{2002MNRAS.329..897H} to trigger the AIC. I assume the mass transfer model used in this work, is based on the Roche model and  ignores the secondary's spin angular momentum. The angular momentum loss due to the  gravitational radiation is calculated. 
This is valid for the WD primary \citep {2001MNRAS.327..888R}. This will further change the orbital period and the eccentricity \citep{2010ChPhL..27k9801W}. I will deal in this section with the binding energy of
a dynamical system based on the change in orbital energy. This could occur during the pre-SNe orbit and immediately before  the explosion, as described by the equation

\textbf{\begin{equation}
 E_{orb,i} =  \frac{- GM_{WD} M_{c,i}}{2a_{i}}
 \end{equation}}

    \textbf{\begin{equation}
 = - \frac{GM_{WD} M_{c,i}}{a_{i}}  +  \frac{M_{WD}M_{c,i}}{M_{NS}+M_{c,i}} v^{2}_{i}
 \end{equation}}

where the $v_{i}$ is WD's initial velocity relative to its companion
 (mass= Chandrasekhar mass  $M_{ch}$).  M$_{NS}$ is the  NS's mass. G is Newton’s
gravitational constant. At the evolution of a binary under instantaneous mass loss, the initial mass of the companion $M_c,i$ is assumed to be at least $4M_{\odot}\leq M_{com} \leq6M_{\odot}$, which may be reasonable for AIC binaries where the companion is on the main sequence of helium. $a_{i}$ is the system's initial semi-major axis, which is approximately (1.5 R$_{\odot}$). For more information, \citep[see e.g.,][]{1983AJ.....88.1857H} and references therein.

The orbital energy after the SNe

\textbf{\begin{equation}
 E_{orb,f} =  \frac{- GM_{NS} M_{c,f}}{2a_{f}}
 \end{equation}}

    \textbf{\begin{equation}
 = - \frac{GM_{NS} M_{c,f}}{a_{f}}  +  \frac{M_{NS}M_{c,f}}{M_{NS}+M_{c,f}} v^{2}_{f}
 \end{equation}}
 
  where the companion $M_{c,f}$'s final mass is estimated to be at least 1.4M$_{\odot}$ [see, for example \citep[see e.g.,][]{2006Tauris}.   The final velocity of WD relative to its companion is given by $v_{f}$. a$_{f}$ is the system's final semi-major axis after the AIC process, which is approximately  (3R$_{\odot}$) (see i.e, \cite{10.1093/mnras/staa3701, 2020arXiv200203011T}. The geometry of an asymmetric SNe in the binary system is illustrated in Fig. ~\ref{fig2}. This will depend on the mass loss and angular momentum variations. The low binding energy of the ejecta and the low explosive energy are, of course, related. The slow rotation is caused by the accretion of gas with small angular momentum during this process.

It is worth noting that different amounts of deposition energy into the common envelope should make a noticeable difference in the two different phases (pre- and post AIC). This is due to the dissipating mechanism acts on the dynamical orbital evolution. In addition, I also consider the angular momentum that could affect  the  separation distance for the system based on the amount of accreted mass (see i.e, \cite{10.1093/mnras/staa3701}.
 
\begin{equation}
    \textit{J}= \frac{M_{NS} M_{c,f}}{M_{NS}+ M_{c,f}} a_{f} \frac{2\pi}{P}
 \end{equation}
 
However, this effect could  be observed from the gravitational radiation emitted by the binary motion in a circular orbit \citep{2004ApJ...614..914A}. In addition, the role of the compactness C = GM/(c$^{2}$R), where c is the speed of light and G is the gravitational constant, during accretion has been adopted from \cite{2021arXiv210313605H}, and points out the need for of constraining the equation of state. As a result, more accreted baryonic mass will be transformed into binding energy instead of gravitational mass \citep{2021ApJ...910L..22H}.

Here, I consider the pre-AIC and post-AIC binding energies of the binary system by assuming the difference in final and initial  orbital energies ($E_{orb,f}$ and $E_{orb,i}$) respectively (see Fig. ~\ref{fig3}). This could cause the system's orbit to continually circularize and contract due to losing orbital energy. 
\begin{equation}
\Delta E_{orb}= E_{orb,f} - E_{orb,i}
 \end{equation}

Following the \cite{1998AJ....116.1009R}, I employ  the efficiency parameter ($\zeta$)  during a common envelope evolution \citep {1976IAUS...73...75P} to obtain the  binding energy, which is dependent on the detailed structure (i.e. radius of the companion  at an evolutionary stage) of the dynamical process in binaries \citep{2000A&A...360.1043D}. Thus, I have

\begin{equation}
 \zeta = \frac{\Delta E_{orb}}{\Delta E_{binding}}
 \end{equation}

From this, I  calculate the $a_{f}$ of the
new binary after the AIC process (see Fig. ~\ref{fig4}). 
If the value of $a_{f}$ is too small, the core will overfill its new Roche lobe \citep{2015MNRAS.447.1713B}, then the cores will merge. As a result,  gravitational waves may be produced during the pre-SNe phase \citep{2017ApJ...846..170T}. 

I  also deal with a system that orbits around a common center of mass \citep[see e.g.,][]{2002MNRAS.329..897H, 2022MNRAS.517.3993M}. Then, $a_{f}$ grows, but the system remains bounded by $a = \frac{r(1+e~cos\theta)}{1-e^{2}}$.
As a result, I can get both
$\Delta E_{orb,i}$ and $ E_{orb,f}$.

The initial binding energy of the system can be
\begin{equation}
E_{binding,i} \simeq {G~M_{b}} (\frac{1}{2a_{f}} - \frac{1}{2a_{i}} )
\end{equation}

where $M_{b} = (M_{WD}+M_{c,i}) = (M_{NS}+M_{c,f}$)

Hence, the final binding energy could be as

\begin{equation}
 E_{binding,f}= E_{binding,i} + \frac{\Delta E_{orb}}{\zeta}
 \end{equation}


In the picture presented here, I find that the initial binding energy will have a significant effect on the final fate of both  evolution of the components (see Fig. ~\ref{fig5}), as well as on the orbital parameters that change during the conservative mass transfer. The majority of the mass is lost in an AIC binary system comes from the WD converting baryonic mass into binding energy during the collapse  of the newborn NS, with a mass change of the order of $0.18M_{\odot}$ \citep{2022JHEAp..35...83T}. This can be happened due to their larger compactness \cite{2021arXiv210313605H}. If  enough orbital energy  continues to be lost, this would cause a merger of the binary components (like in some CO WDs merging). At this stage of evolution, this phenomenon created a common envelope phase by reducing orbital energy by envelope binding energy \citep{2004ApJ...612.1044P, 2015MNRAS.447.1713B} (see Fig. ~\ref{fig6}). 

A note should be made concerning the given equation of state, because it will govern the specific binding energy for determining the binary's evolution. \cite{PhysRevD.102.103011} found a strong correlation between  
several equations of state and tidal deformability associated with NS binding energies. Furthermore, they discovered that the lower limit of NS mass has a binding energy of 1.52$\times10^{53}$ ergs.  Based on these interesting results, \cite{2021MNRAS.506.1462N}
 estimated the amount of the gravitational mass of the NS, which is $ \rm \sim 1.2M_{\odot}$. This result is consistent with the NS observations made by \cite{2016ARA&A..54..401O, 2022PASA...39...40T}.

The result from the mass estimation benefits from the mildly recycled pulsar J0453+1559 \citep{2015ApJ...812..143M}. This system consists of a double NS with masses of $1.56M_{\odot}$  and $1.17M_{\odot}$, an orbital period of 4.07-day, a spin period of 45 ms, and an orbital eccentricity of (e =0.11).  The amount of gravitational binding energy is expected to be around ($\Delta M \rm \sim 0.18M_{\odot}$) as shown in Fig. ~\ref{fig6}.  This can be used to add some additional constraints when given a gravitational mass measurement for double NSs.





\section{Conclusions}

An interesting point should be made about  the outcome of the final binding energy within the framework of the AIC process, taking into consideration the non-zero velocity of ejected matter. However, the equation of state influences how much binding energy  an NS has. As a result, the larger  the binding energy, the more compact  the NS. The binding energy, thus dominates the gravitational waves at a given amount of accreted mass associated with the angular momentum loss. This would provide more insight into the AIC observations in the future.  I have also investigated the effect of accretion on the orbital period on a NS's binding energy. With the use of the orbital period along with a donor star mass evolution, I have shown that it is possible to add some constraints on the expected values of both mass and  binding for NS energy parameters that depend on the accretion rate. However, this  will depend on the initial binding energy associated with the two systems' initial orbital energies, whether an envelope is present or not, since the final binding energy will depend on the initial binding energy. 
This would help us   figure out the accretion-driven evolution modes for binaries.
 The corresponding increase in orbital energy propels the binary into a longer-period orbit.
 This causes it to collapse and violently release its
gravitational energy, which might be observable by  gravitational
wave observatories such as  the LIGO-Virgo
Collaboration.

\section*{Acknowledgments}
Part of the content of this manuscript has been presented at the IAU Symposium 366, [doi:10.5281/zenodo.5759007]. Special thanks to Nour AlMusleh for helping the calculations. 
The author would  like to thank  the anonymous referees for the careful reading of the manuscript and for all suggestions and comments which allowed us to improve both the quality and the clarity of the paper.

\bibliography{Template}%
%

\end{document}